\documentclass[pra,showpacs,showkeys]{revtex4}% Physical Review A

\usepackage{amsmath}
\usepackage{amsfonts}            
\usepackage{amssymb}

\usepackage{mathrsfs}

\usepackage{eufrak}

\usepackage{graphicx}% Include figure files
\usepackage{epstopdf}% Include figure files

\usepackage{dcolumn}% Align table columns on decimal point
\usepackage{bm}% bold math 

\makeatletter
\def\@dotsep{4.5}
\makeatother

\begin{document}

\title{Supersymmetry identifies molecular Stark states \\ whose eigenproperties can be obtained analytically}

\author{Mikhail Lemeshko}

\email{mikhail.lemeshko@gmail.com}

\affiliation{%
Fritz-Haber-Institut der Max-Planck-Gesellschaft, Faradayweg 4-6, D-14195 Berlin, Germany
}%

\author{Mustafa Mustafa}

\email{mustafa@purdue.edu}

\affiliation{%
Department of Chemistry and Physics and Birck Nanotechnology Center, Purdue University, West Lafayette, Indiana 47907, USA
}%

\author{Sabre Kais}

\email{kais@purdue.edu}

\affiliation{%
Department of Chemistry and Physics and Birck Nanotechnology Center, Purdue University, West Lafayette, Indiana 47907, USA
}%

\author{Bretislav Friedrich}

\email{brich@fhi-berlin.mpg.de}

\affiliation{%
Fritz-Haber-Institut der Max-Planck-Gesellschaft, Faradayweg 4-6, D-14195 Berlin, Germany~}%

\date{\today}% It is always \today, today,

\begin{abstract}

We made use of supersymmetric (SUSY) quantum mechanics to find a condition under which the Stark effect problem for a polar and polarizable closed-shell diatomic molecule subject to collinear electrostatic and nonresonant radiative fields becomes exactly solvable. The condition, $\Delta \omega = \frac{\omega^2}{4 (m+1)^2 }$,  connects values of the dimensionless parameters $\omega$ and $\Delta \omega$ that characterize the strengths of the permanent and induced dipole interactions of the molecule with the respective fields. The exact solutions are obtained for the  $|\tilde{J}=m,m;\omega,\Delta \omega\rangle$ family of ``stretched'' states.  The field-free and strong-field limits of the combined-fields problem were found to  exhibit supersymmetry and shape-invariance, which is indeed the reason why they are analytically solvable. By making use of the analytic form of the $|\tilde{J}=m,m;\omega,\Delta \omega\rangle$  wavefunctions, we obtained simple formulae for the expectation values of the space-fixed electric dipole moment, the alignment cosine, the angular momentum squared, and derived a ``sum rule'' which combines the above expectation values into a formula for the eigenenergy. The analytic expressions for
the characteristics of the strongly oriented and aligned
states provide a direct access to the values of the interaction
parameters required for creating such states in the
laboratory.

\end{abstract}

 %Use showkeys class option if keyword

\maketitle

\section{Introduction}

Whether a problem in quantum mechanics is exactly solvable is closely related to its supersymmetry (SUSY)~\cite{DuttKhareSukhatmeAJP88, CooperPhysRep95}. The foundations of supersymmetric quantum mechanics have been worked out by Witten in 1981 as an example of SUSY in zero-dimensional field theory~\cite{WittenNuclPhysB81}. Soon thereafter, supersymmetric quantum mechanics rapidly evolved into a new branch of mathematical physics~\cite{DuttKhareSukhatmeAJP88, CooperPhysRep95, SatoTanaka2002} and reached a highpoint in 1983 -- then unnoticed -- when Gendenshtein established a connection between supersymmetry and exact solvability~\cite{Gendenshtein83}: he demonstrated that Schr\"odinger's equation is exactly solvable
if the potential and its superpartner exhibit shape-invariance. Whereas a supersymmetric Hamiltonian can be constructed for any potential whose ground-state wavefunction is analytic, shape invariance only exists for supersymmetric potentials that are interconvertible by a change of a parameter other than the integration variable itself~\cite{KaisJCP89}. Herein, we make use of the methods of supersymmetric quantum mechanics to arrive at exact wavefunctions and other eigenproperties of linear molecules subject to nonresonant electric fields  in closed form.

In our previous work on the molecular Stark effect, we showed that for polar molecules, combined collinear electric and nonresonant radiative fields can synergetically produce spatially oriented pendular states, in which the molecular axis librates over a limited angular range about the common field direction~\cite{HaerteltFriedrichJCP08}. These directional states comprise coherent superpositions or hybrids of the field-free rotational states $|J,m\rangle$, with a range of $J$ values but a fixed value of $m$, which remains a good quantum number by virtue of the azimuthal symmetry about the fields. This has proved an effective and versatile means to produce oriented molecules for applications ranging from molecule optics and spectroscopy to chemistry and surface science~\cite{StapelfeldtSeidemanRMP03, KremsPCCP08, VattuoneProgSurfSci10}. However, the eigenproperties of the Stark states in question had to be found numerically, typically by diagonalizing a truncated Hamiltonian matrix. Here we show that supersymmetric factorization of the Hamiltonian yields exact wavefunctions $|\tilde{J}=m,m;\omega,\Delta \omega\rangle$ in closed form for a particular ratio of the parameters $\omega$ and $\Delta \omega$ that determine the interaction strengths of the molecules with the static and radiative fields, respectively. 

The paper is organized as follows. In Section~\ref{sec:general}, we first introduce the molecular Stark effect problem for the case of collinear electrostatic and radiative fields.
In Section~\ref{sec:SUSYfactor} we use methods of supersymmetric quantum mechanics to find solutions to the problem of molecules in combined fields in closed form. We derive a relation between strengths of the electrostatic and laser fields, $\omega$ and $\Delta \omega$, at which the wavefunctions of the $|\tilde{J}=m,m;\omega,\Delta \omega\rangle$ states take a simple analytic form and show that SUSY furnishes isospectral partner potentials that can be realized with combined fields. In Section~\ref{sec:SIP} we discuss shape-invariance of the supersymmetric partner potentials obtained and the general conditions for the exact solvability of the Stark effect problem.  In Section~\ref{sec:Limits} we investigate the field-free and strong-field limits, where the field strengths $\omega$ and $\Delta \omega$ approach, respectively, zero or infinity. We show that in both limits, the problem exhibits supersymmetry and shape-invariance and, therefore, can be solved exactly. Section~\ref{sec:Applications} gives a summary of the exact closed-form expressions for the properties of molecules in fields obtained from the exact wavefunctions. These properties include the eigenenergy, the space-fixed dipole moment, the alignment cosine, and the expectation value of angular momentum. The availability of these otherwise hard-to-come-by properties in closed form allows to reverse-engineer the problem of finding the interaction parameters required for creating quantum states with preordained characteristics. The main conclusions of the work are summarized in Section~\ref{sec:Conclusions}. An appendix surveys the concepts pertinent to exact solvability.

\section{Molecules in collinear electric and radiative fields}
\label{sec:general}

We consider a $^1\Sigma$ linear molecule with a rotational constant $B$, a permanent dipole moment $\mu$ along the internuclear axis, and polarizability components $\alpha_\parallel$ and $\alpha_\perp$ parallel and perpendicular to the internuclear axis. The molecule is subjected to an electrostatic field $\boldsymbol{\varepsilon}$ combined with a nonresonant laser field of intensity $I$, whose linear polarization is collinear with $\boldsymbol{\varepsilon}$. With energy expressed in terms of $B$, the Hamiltonian takes the dimensionless form~\cite{FriHerJCP99}:
\begin{equation}
	\label{HcombinedFields}
	H =  \mathbf{J}^2  + V_{\mu, \alpha} (\theta),
\end{equation}
with the angular momentum operator
\begin{equation}
	\label{j2explicit}
	\mathbf{J}^2 = - \frac{1}{\sin \theta} \frac{\partial}{\partial \theta} \left( \sin \theta  \frac{\partial}{\partial \theta}  \right) - \frac{1}{\sin^2 \theta} \frac{\partial^2}{\partial \phi^2}
\end{equation}
and the interaction potential
\begin{equation}
	\label{Vcombined}
	V_{\mu, \alpha} (\theta)  = - \omega \cos \theta - \left( \Delta\omega \cos^2 \theta + \omega_\perp   \right)
\end{equation}
The dimensionless interaction parameters are given as $\omega \equiv \mu \varepsilon/B$ and  $\Delta \omega \equiv \omega _{||}-\omega _{\bot }$, with $\omega _{||,\bot } \equiv 2\pi \alpha_{||,\bot }I/(Bc)$.

The common direction of the collinear electrostatic and linearly polarized radiative fields defines an axis of cylindrical symmetry, chosen to be the space-fixed axis $Z$. The projection, $m$, of the angular momentum $\mathbf{J}$ on $Z$ is then a good quantum number while $J$ is not. However, one can use the value of $J$ of the field-free rotational state, $Y_{J, m} (\theta, \phi)$, that adiabatically correlates with the hybrid state as a label, designated by $\tilde{J}$, so that $\psi_{\tilde{J}, m}^{\omega, \Delta \omega} (\theta, \phi) \to Y_{J, m} (\theta, \phi) $ for $\omega, \Delta \omega \to 0$. For arbitrary interaction strengths, the solution to the the Schr\"odinger equation with Hamiltonian~(\ref{HcombinedFields}) is an infinite 
coherent superpositions of the field-free wavefunctions,
\begin{equation}
	\label{PendularState}
	|\tilde{J}, m;\omega, \Delta \omega \rangle = \sum_{J} c_{J m}^{\tilde{J}, m} (\omega, \Delta \omega) Y_{J m} ,
\end{equation}
whose expansion coefficients $c_{J m}^{\tilde{J}, m} (\omega, \Delta \omega)$ can be obtained by truncating series~(\ref{PendularState}) at some maximum value of $J$ and then diagonalizing Hamiltonian~(\ref{HcombinedFields}) in the finite basis set of the field-free wavefunctions. 

The axial symmetry of the problem allows to separate angular variables and express the dependence on the azimuthal angle $\phi$ via the good quantum number $m$. In this case the Schr\"odinger equation with Hamiltonian~(\ref{HcombinedFields}) can be recast as
\begin{equation}
	\label{SEcombinedfields}
	\left[ - \frac{1}{\sin\theta} \frac{d}{d \theta} \left( \sin\theta  \frac{d}{d \theta}  \right)  + V_{\mu, \alpha}^\text{(3D)} (\theta)   \right]   \psi (\theta) = E \psi (\theta),
\end{equation}
where the three-dimensional effective potential is given by
\begin{equation}
	\label{V3D}
	V_{\mu, \alpha}^\text{(3D)} (\theta)  = \frac{m^2}{\sin^2 \theta} - \omega \cos \theta - \Delta \omega \cos^2 \theta
\end{equation}
We note that all rotational levels are uniformly shifted by $\omega_\perp$. In what follows we use  $E = E_{\mu,\alpha} + \omega_\perp$ instead of the `true' molecular energy, $E_{\mu,\alpha}$. Moreover, since the Stark effect does not depend on the sign of $m$, we can define the projection of the angular momentum on $Z$ as a positive quantity, $m \equiv |m|$. 

By means of the substitution~\cite{InfeldHullRMP51},
\begin{equation}
	\label{PsiToFsubst}
	\psi (\theta) = f (\theta) (\sin  \theta)^{-\frac{1}{2}},
\end{equation}
Schr\"odinger equation~(\ref{SEcombinedfields}) can be transformed to a one-dimensional form,
%\begin{equation}
%	\label{SEspher}
%	\left [  - \frac{1}{\sin \theta} \frac{d}{d\theta} \left( \sin\theta \frac{d}{d\theta} \right)  - 2\beta\cos\theta - \beta^2 \cos^2\theta + \beta^2  \right ] \psi_0(\theta) = 0
%\end{equation}
%\begin{equation}
%	\label{SEtransformed}
%	H_- f_0(\theta) \equiv \Biggl [ -\frac{d^2}{d \theta^2} + V_- (\theta)   \Biggr] f_0(\theta) = 0,
%\end{equation}
\begin{equation}
	\label{SEtransformed}
	\Biggl [ -\frac{d^2}{d \theta^2} + V_{\mu, \alpha}^\text{(1D)} (\theta)   \Biggr ] f (\theta) = 0,
\end{equation}
where
\begin{equation}
	\label{Veff1D}
	V_{\mu, \alpha}^\text{(1D)} (\theta) = \frac{m^2 - \frac{1}{4}}{\sin^2\theta} - \omega \cos\theta - \Delta \omega \cos^2\theta  - \frac{1}{4} - E
\end{equation}
is a one-dimensional effective potential,
which will be shown to play the role of one of the requisite superpartner potentials leading to ground state energy $E=E_0$.

\section{Supersymmetric factorization of the combined-fields Hamiltonian}
\label{sec:SUSYfactor}

Here we invoke supersymmetry to find analytic solutions to eq. (\ref{SEtransformed}) and subsequently to eq. (\ref{SEcombinedfields}). Supersymmetry makes use of the first-order differential operators, $A^\pm \equiv \mp \frac{d}{d\theta} + W(\theta)$, with $W(\theta)$ the superpotential. The superpartner Hamiltonians are defined by
\begin{equation}
	\label{HminusFactor}
	H_\mp = A^\pm A^\mp =  -\frac{d^2}{d\theta^2} +  V_\mp^\text{(1D)}(\theta),
\end{equation}
with the one-dimensional partner potentials $V_\pm^\text{(1D)} (\theta) \equiv W^2(\theta) \pm W'(\theta)$. The superpartner Hamiltonians have the same energy spectra except for the ground state, and, if the eigenfunctions of one of the partner Hamiltonians $H_\mp$ are known, the eigenfunctions of the other can be obtained analytically via the intertwining relations~\cite{CooperPhysRep95, SukumarJPA85}.

For the case of a molecule in combined fields we start from the following Ansatz for the superpotential:
\begin{equation}
	\label{Wansatz}
	W(\theta) = a \cot (\theta) + q(\theta),
\end{equation}
where the constant $a$ and the function $q(\theta)$ are to be determined. The first term in eq.~(\ref{Wansatz}) is the superpotential for a field-free
rigid rotor, which is a special case of the Rosen-Morse I potential~\cite{CooperPhysRep95}.
The superpotential $W(\theta)$ yields the supersymmetric partner potentials  $V_\pm^\text{(1D)} (\theta)$:
\begin{equation}
	\label{VpmviaQ}
 	V_\pm^\text{(1D)} (\theta) \equiv W^2(\theta) \pm W'(\theta) = \frac{a(a \mp 1)}{\sin^2\theta} +q^2(\theta) \pm q'(\theta) + 2 a q(\theta) \cot\theta  - a^2.
\end{equation}
By identifying the effective potential~(\ref{Veff1D}) with $V_-^\text{(1D)}(\theta)$ and substituting it into eq.~(\ref{VpmviaQ}), we obtain
\begin{equation}
	\label{ComparingPots}
 	\frac{m^2 - \frac{1}{4}}{\sin^2\theta} - \omega \cos\theta - \Delta \omega \cos^2\theta  -\frac{1}{4} - E_0  = \frac{a(a+1)}{\sin^2\theta} +q^2(\theta) - q'(\theta) + 2 a q(\theta) \cot\theta  - a^2,
\end{equation}
which is satisfied for 
\begin{equation}
	\label{qi}
	q(\theta) = \beta \sin \theta,
\end{equation}
\begin{equation}
	\label{ai}
	a=-(m+1/2),
\end{equation}
\begin{equation}
	\label{EnergPsi}
	E_0  = m(m+1) - \beta^2,
\end{equation}
and
\begin{equation}
	\label{OmegaDeltaOmega}
	\Delta \omega = \frac{\omega^2}{4 (m+1)^2 } = \beta^2.
\end{equation}
Substitution from eqs. (\ref{qi}) and (\ref{ai}) into eq.~(\ref{Wansatz}) leads to a superpotential
\begin{equation}
	\label{W}
	W(\theta) =  - \left( m + \frac{1}{2} \right) \cot \theta + \beta \sin \theta,
\end{equation}
corresponding to the following pair of superpartner potentials:
\begin{equation}
	\label{Vminus1D}
	V_-^\text{(1D)}(\theta) =   \frac{m^2 - \frac{1}{4}}{\sin^2 \theta} - 2\beta(m+1) \cos\theta - \beta^2\cos^2\theta + \beta^2 - m(m+1) -\frac{1}{4} 
\end{equation}
\begin{equation}
	\label{Vplus1D}
	V_+^\text{(1D)}(\theta) =    \frac{(m+1)^2 - \frac{1}{4}}{\sin^2 \theta} - 2\beta m \cos\theta - \beta^2\cos^2\theta + \beta^2 - m(m+1) -\frac{1}{4}
\end{equation}
We note that by identifying the effective potential~(\ref{Veff1D}) with $V_+^\text{(1D)}(\theta)$ instead, the result would have been the same for $a= (m-1/2)$.

Thus, for each value of the good quantum number $m$ there exists a pair of the supersymmetric partner Hamiltonians, $H_-$ and $H_+$, which are isospectral and fulfill the following relations:
\begin{align}
	\label{HplusAfmin}
	H_+ (A f_n^-(\theta)) &= E_n^- (A f_n^-(\theta)); \\
	H_- (A^\dag f_n^-(\theta)) &= E_n^+ (A^\dag f_n^-(\theta))
\end{align}
which imply:
\begin{equation}
	\label{Erelation}
	E_n^+ = E_{n+1}^-; \hspace{0.3cm} E_0^- = 0
\end{equation}
\begin{equation}
	\label{PsiplusiaPsiminus}
	f_n^+(\theta) = (E_{n+1}^-)^{-1/2} A f_{n+1}^-(\theta);  \hspace{0.3cm}  f_{n+1}^-(\theta) = (E_{n}^+)^{-1/2} A^\dag f_{n}^+(\theta),
\end{equation}
\begin{equation}
	\label{gr}
A f_0^- (\theta) = 0
\end{equation}

\begin{figure}
\begin{tabular}{ c  c c c }
\includegraphics[width=5cm]{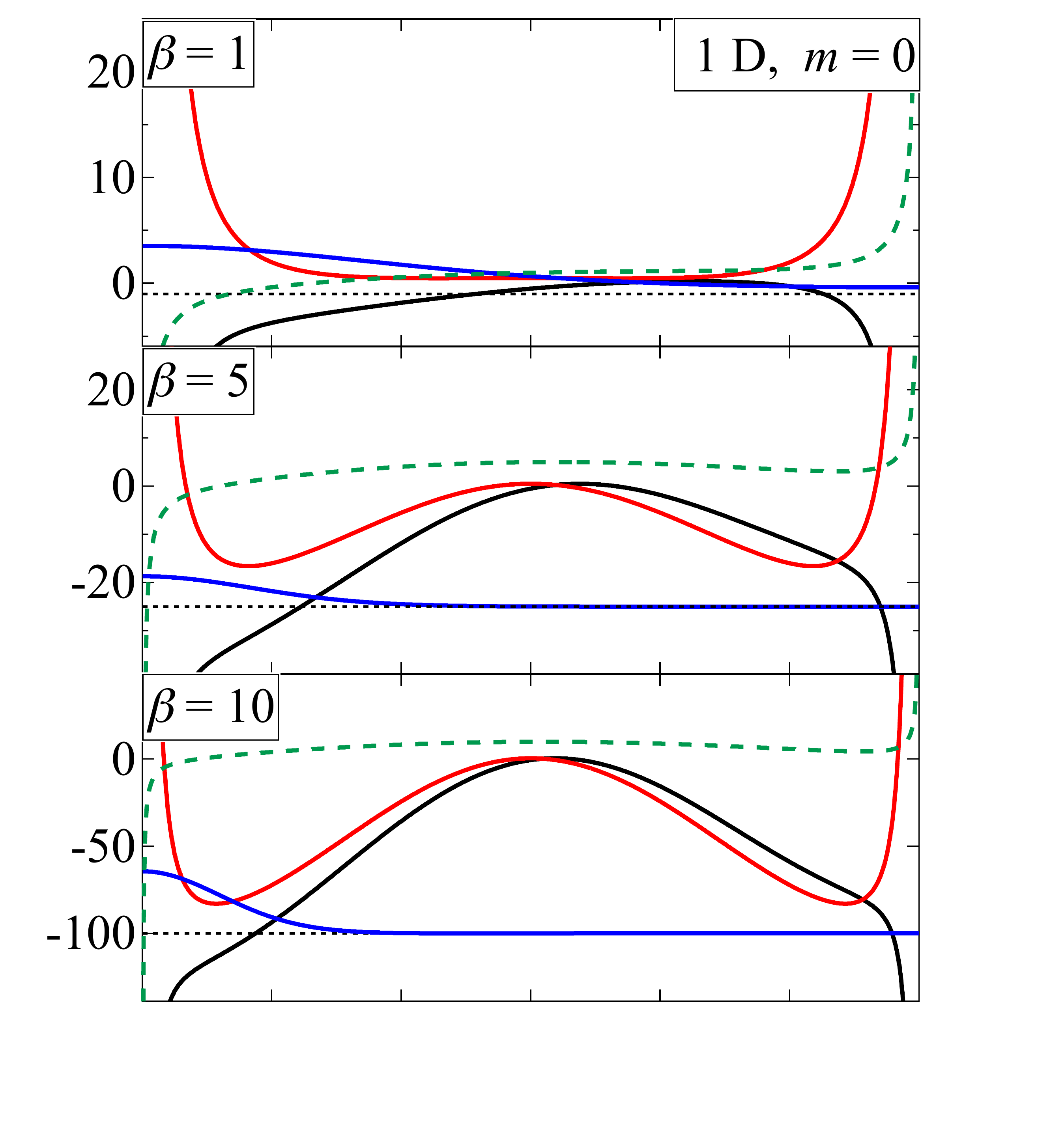} &\hspace{-1cm} \includegraphics[width=5cm]{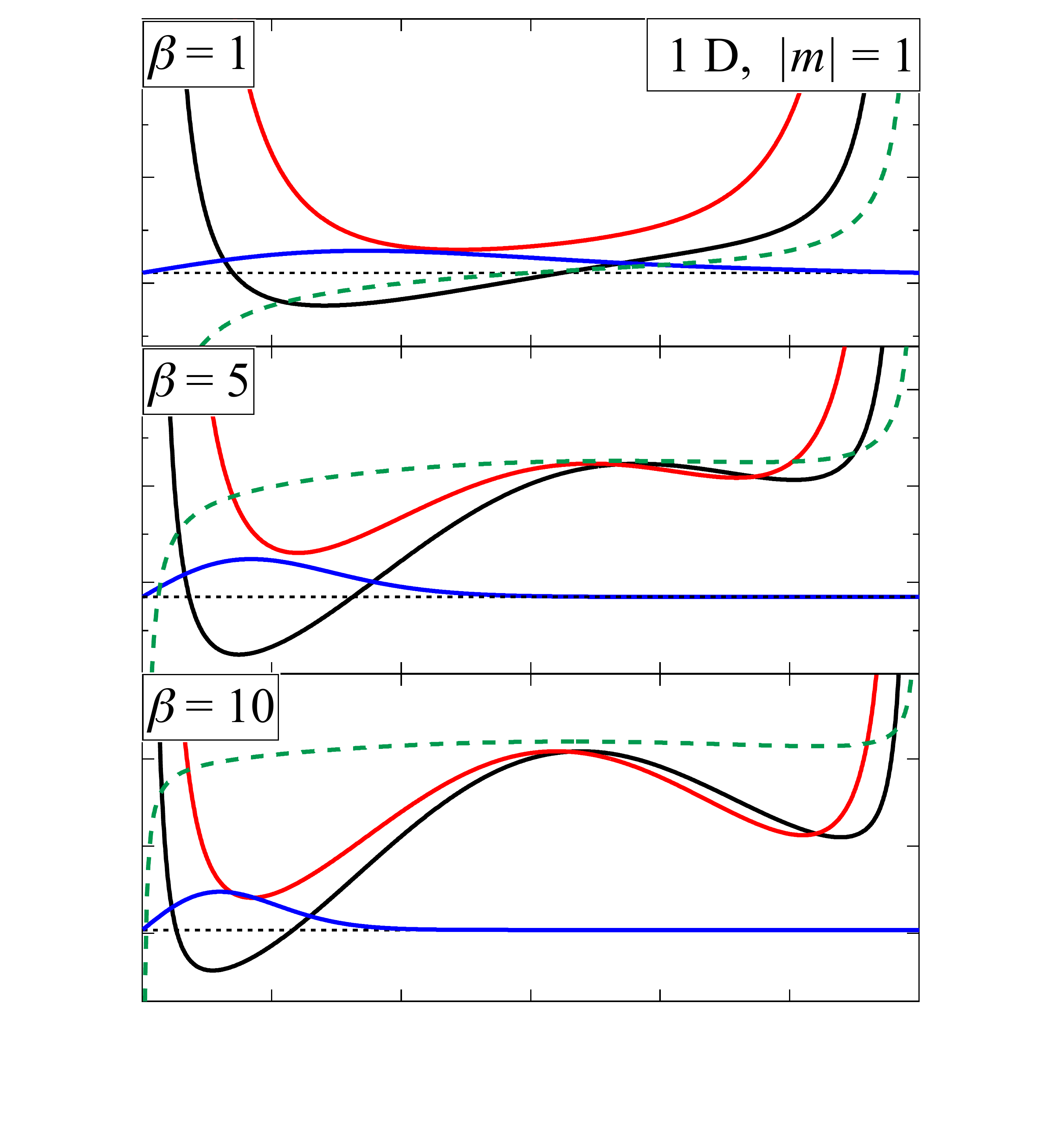} &\hspace{-1cm} \includegraphics[width=5cm]{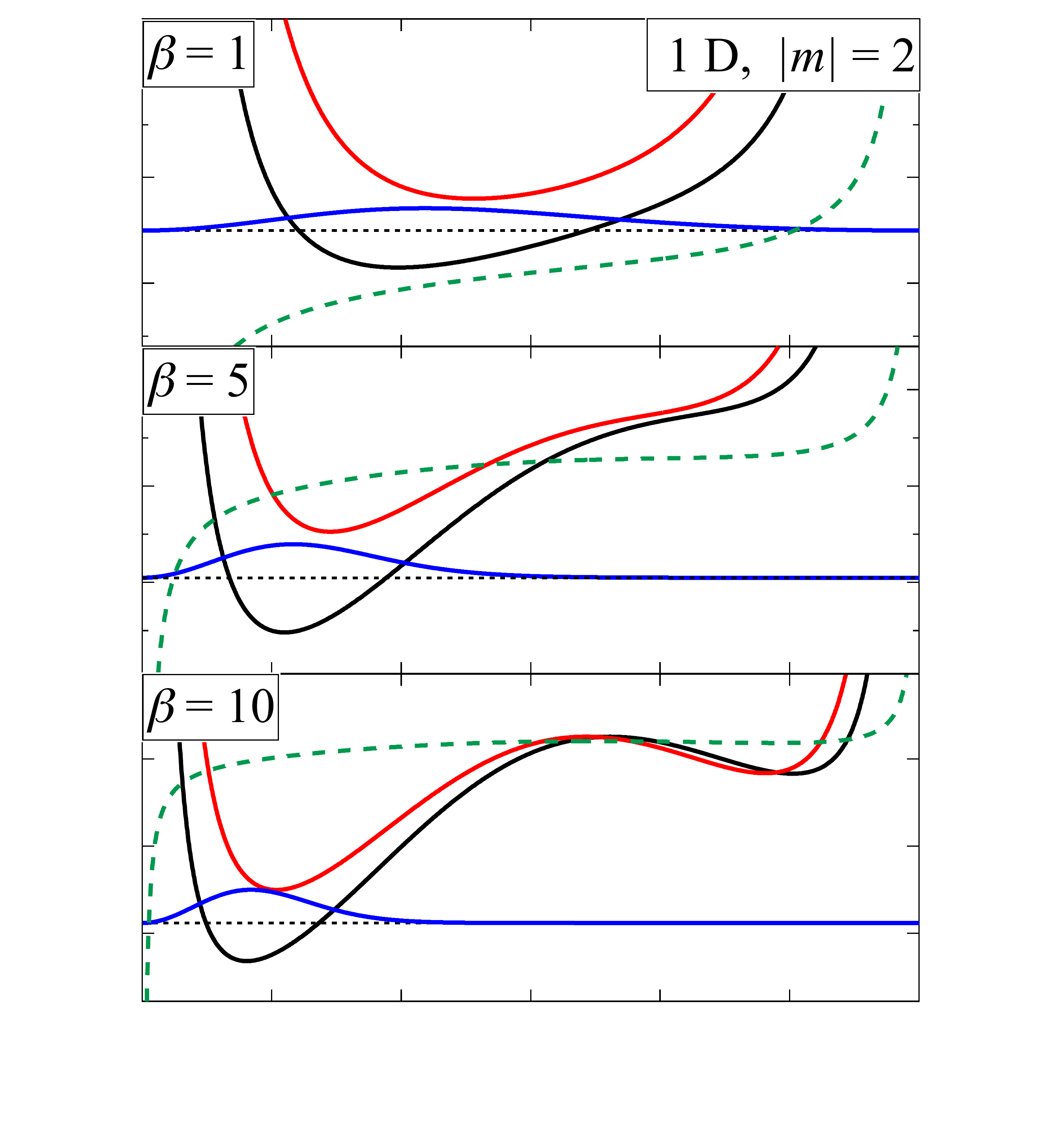} &\hspace{-1cm} \includegraphics[width=5cm]{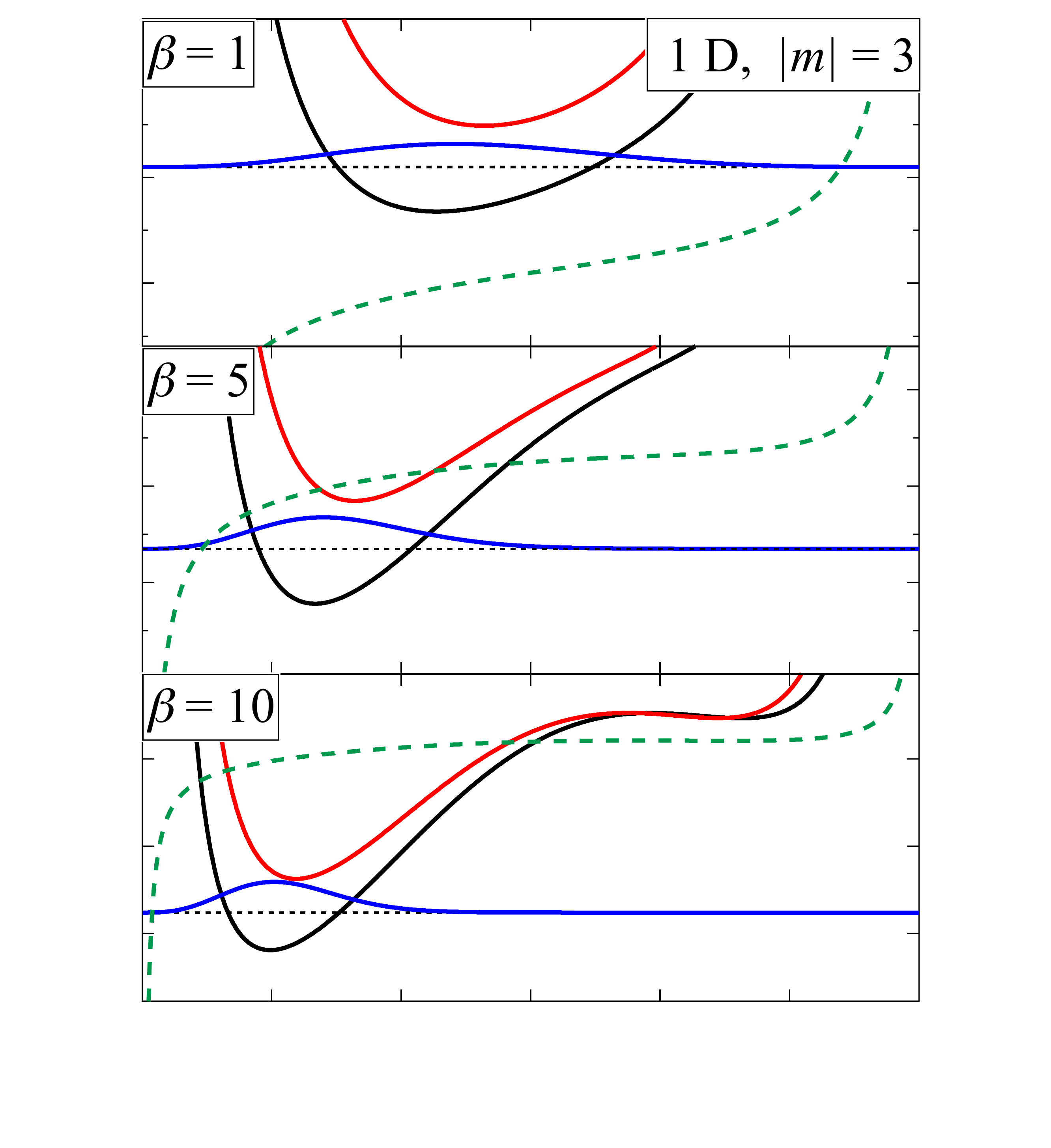} \\[-17pt]
 \includegraphics[width=5cm]{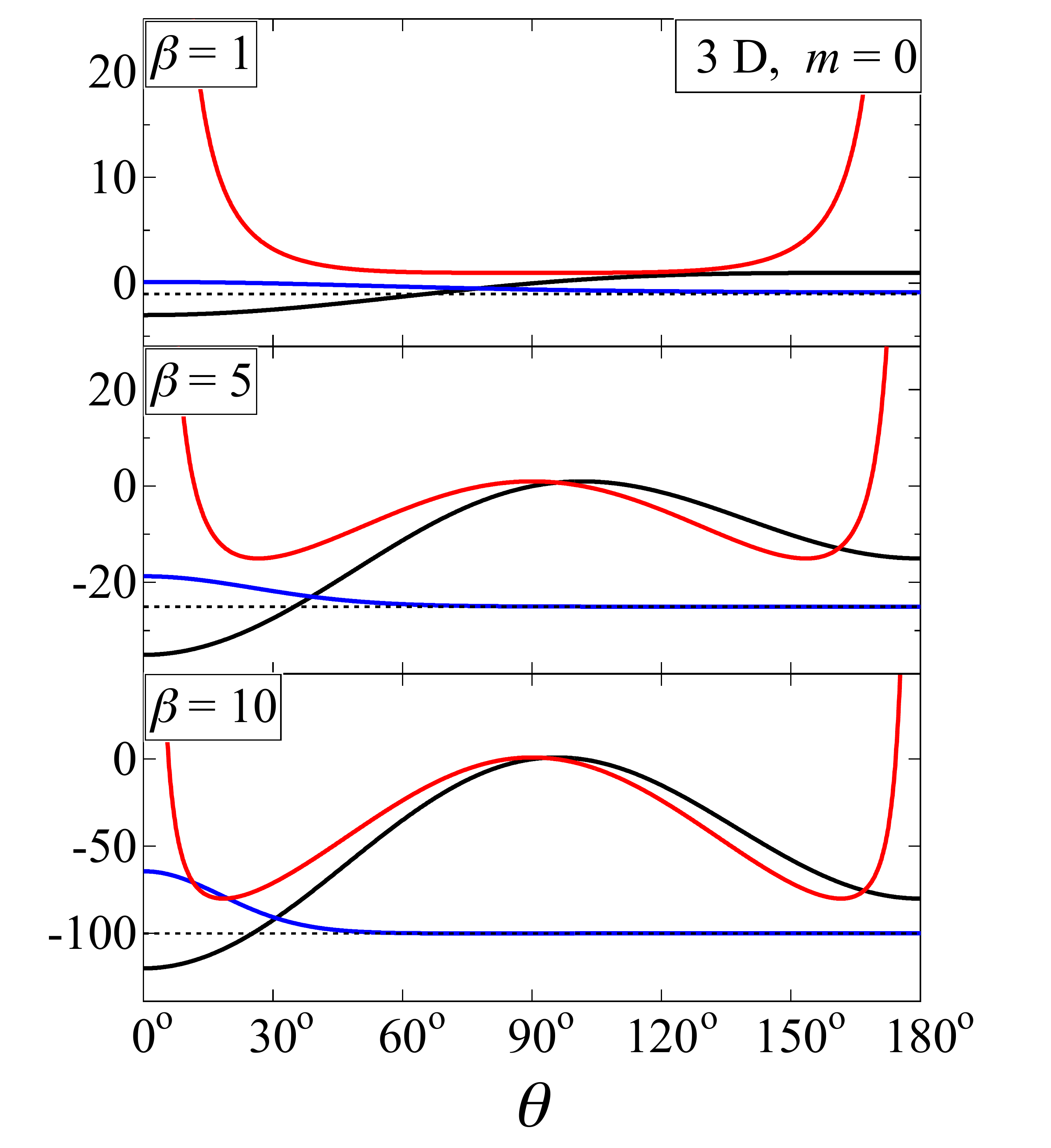} &\hspace{-1cm} \includegraphics[width=5cm]{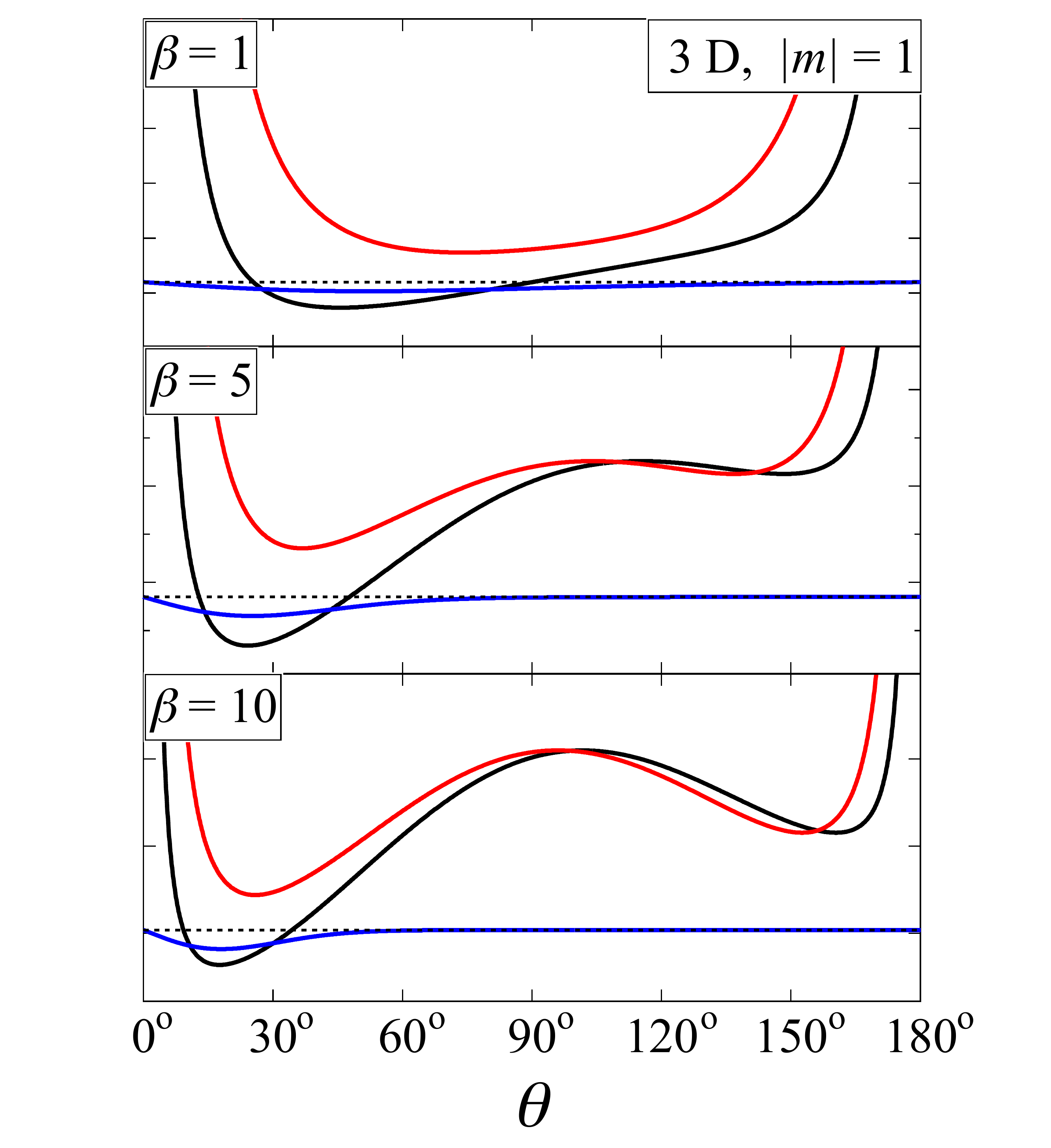} &\hspace{-1cm} \includegraphics[width=5cm]{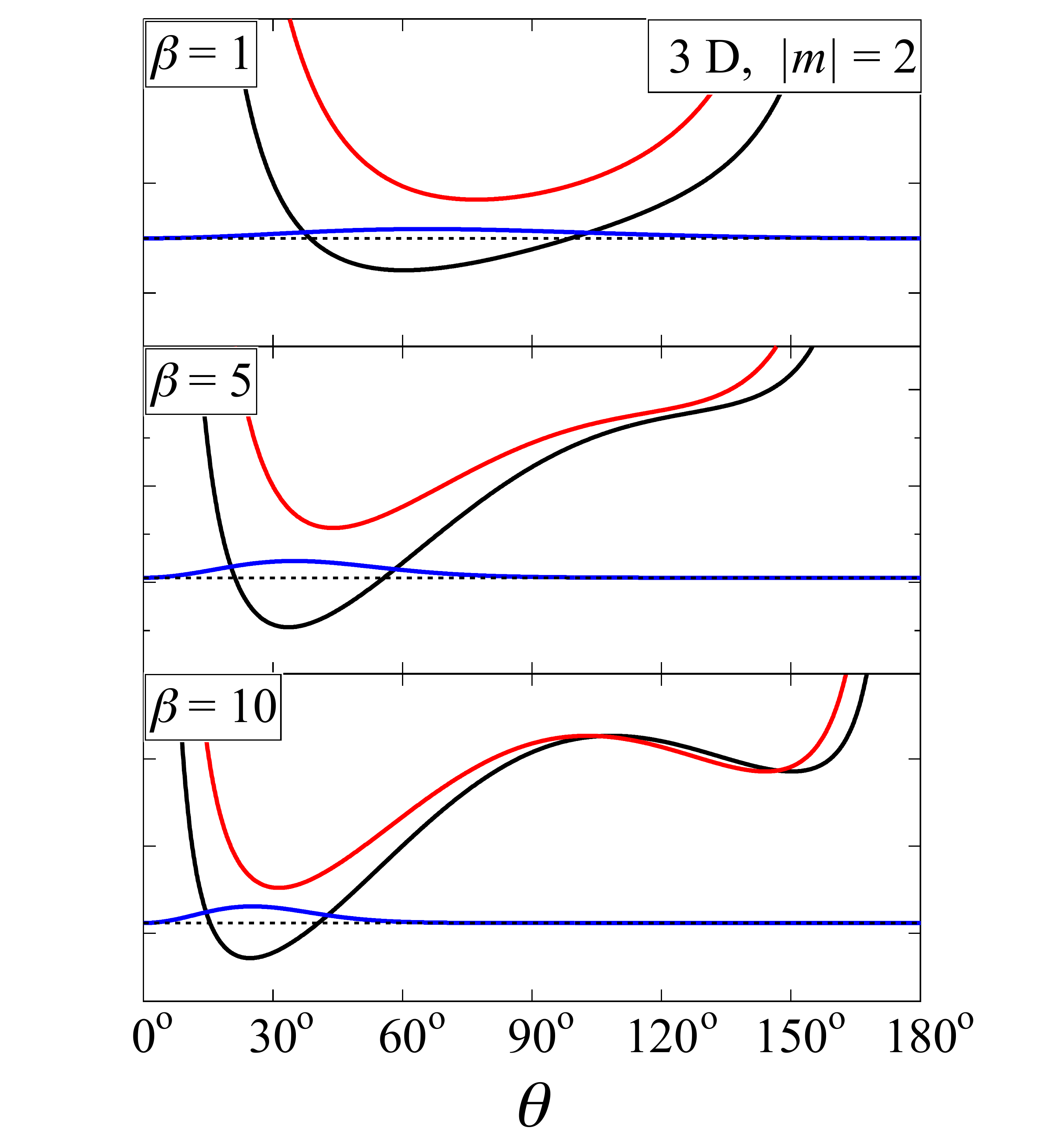} &\hspace{-1cm} \includegraphics[width=5cm]{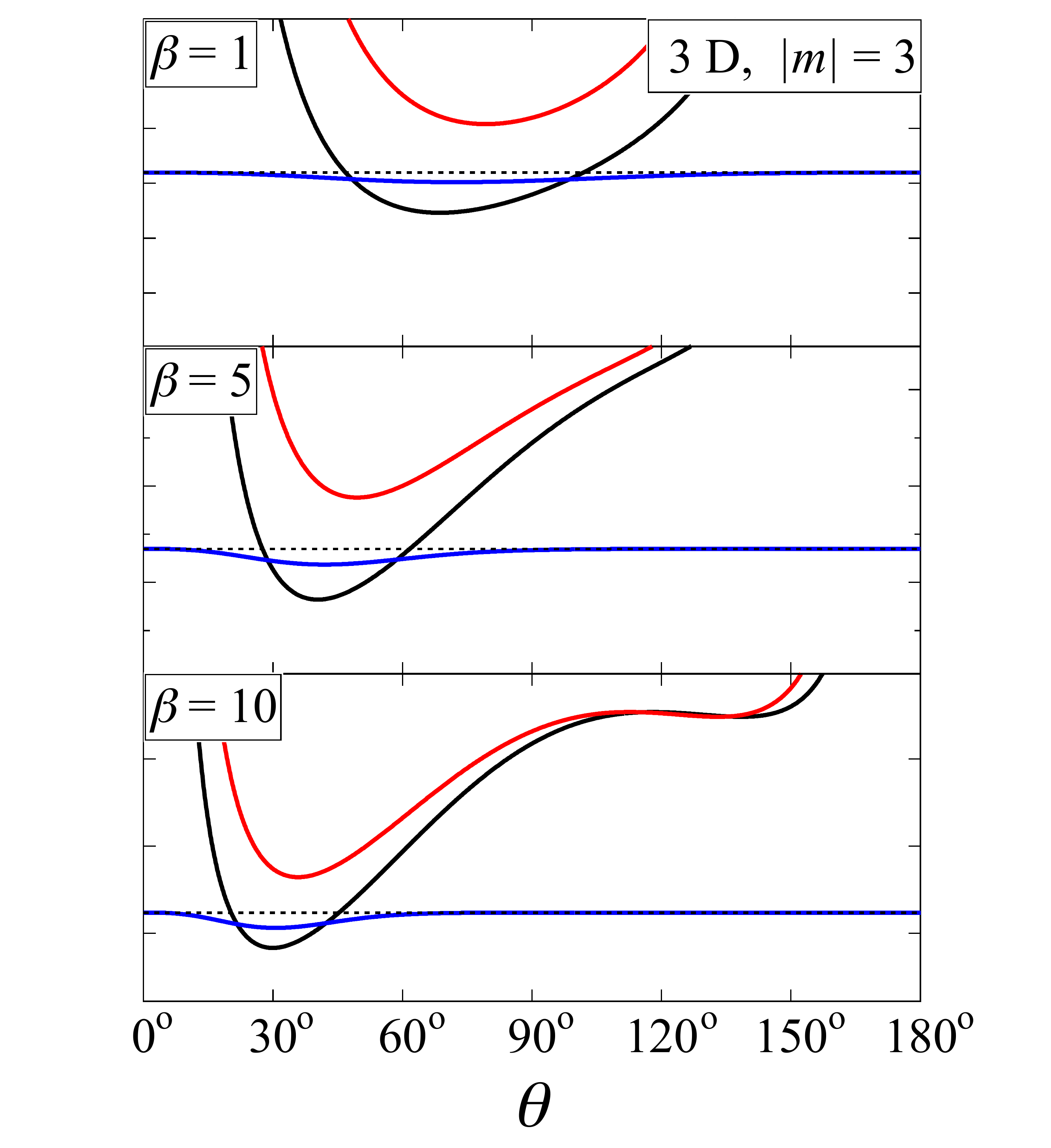} \end{tabular}
\caption{\label{fig:V_minus_plus_m01} Top panels: 1D supersymmetric partner potentials $V_-^\text{(1D)}(\theta)+E_0$ [black solid line], $V_+^\text{(1D)}(\theta)+E_0$ [red solid line], superpotential $W(\theta)$ [green dashed line], and ground-state wavefunction $f_0(\theta)$ [blue solid line] for $|m|=0,1,2,3$ and different values of $\beta$. Bottom panels: 3D supersymmetric partner potentials, $V_-^\text{(3D)}(\theta)$ [black solid line] and $V_+^\text{(3D)}(\theta)$ [red solid line] corresponding to the above 1D potentials, along with the ground-state wavefunction of a molecule in combined fields, $\psi_0(\theta)$ [blue solid line]. The eigenenergy of the ground state, $E_0= m(m+1) - \beta^2$ (in units of $B$), is shown by the black dotted line.}
\end{figure}

The ground state wavefunction $f_0^-(\theta)$ can be obtained from the superpotential (\ref{W}) 
\begin{equation}
	\label{fviaW}
	f_0^-(\theta) = N \exp \left[ - \int_0^\theta W(x) dx \right] 
\end{equation}
which yields the wavefunction in closed form,
\begin{equation}
	\label{fviaW}
	f_0^-(\theta)  = N (\sin\theta)^{(m+1/2)}~e^{\beta \cos \theta}
\end{equation}
Its normalization constant is given by
\begin{equation}
	\label{NormConst}
	N = \left[  \frac{\Gamma (m+3/2) }{2 \pi^{3/2} \Gamma (m+1) {}_0F_1 (; m+3/2; \beta^2)  }  \right ]^{1/2}
\end{equation}
where ${}_0F_1 (; a; z)$ is the confluent hypergeometric function~\cite{AbramowitzStegun}. Since the ground state wavefunction (\ref{fviaW}) is normalizable and obeys the annihilation condition ({\ref{gr}}), the supersymmetry obtained is unbroken~\cite{CooperPhysRep95, WittenNuclPhysB82, CooperFreedmanAnnPhys83}. The top panels of Fig.~\ref{fig:V_minus_plus_m01} show the 1D supersymmetric partner potentials $V_\mp^\text{(1D)} (\theta) + E_0$, the superpotential $W(\theta)$, and the ground-state wavefunction $f_0(\theta)$ for $|m|=0,1,2,3$ and different values of the interaction parameter $\beta$. As $\beta$ increases, the nodeless wavefunction $f_0(\theta)$ becomes strongly confined near the potential minimum.  The potentials (\ref{Vminus1D}) and (\ref{Vplus1D}) come close to each other with growing $m$ and eventually coincide in the limit of $m \gg 1$.

By comparing the 1D potentials $V_+^\text{(1D)}$ and $V_-^\text{(1D)}$, eqs.~(\ref{Vminus1D}),~(\ref{Vplus1D}), with eqs.~(\ref{SEcombinedfields}) and (\ref{SEtransformed}), we obtain the 3D potentials corresponding to a molecule interacting with the combined fields,
\begin{equation}
	\label{Veff3D}
	V_-^\text{(3D)} (\theta) = \frac{m^2}{\sin^2\theta} - 2(m+1) \beta \cos\theta - \beta^2 \cos^2\theta
\end{equation}
\begin{equation}
	\label{Vplus3D}
	V_+^\text{(3D)} (\theta) = \frac{(m+1)^2}{\sin^2 \theta} - 2 m \beta \cos \theta  -  \beta^2 \cos^2\theta
\end{equation}
The ground-state wavefunction of the $V_-^\text{(3D)}$ potential (corresponding to energy $E_0=m(m+1) - \beta^2$) then becomes
\begin{equation}
	\label{psiWF}
	\psi_0(\theta) = N (-1)^m (\sin\theta)^{m}~e^{\beta \cos \theta}.
\end{equation}
The phase factor $(-1)^m$ leads to the correct asymptotic behavior of the wavefunction which, for $\beta = 0$, reduces to the ground state wavefunction of a rigid rotor with $J=m$, $Y_{m,m} (\theta, 0)$.

Eq.~(\ref{Vplus3D}) coincides with the potential for a rigid rotor in the combined fields whose projection quantum number is $m+1$ and whose interaction strengths are related by
\begin{equation}
	\label{OmegaDelatOmega2}
	\Delta \omega = \frac{\omega^2}{4 m^2 } = \beta^2,
\end{equation}

Hence we have shown that, given a value of $\beta$, the Hamiltonian of a molecule with a projection  $m$ of the angular momentum on the combined fields whose interaction parameters are related by eq. (\ref{OmegaDeltaOmega}) has the same set of eigenvalues as a molecule with a projection $m+1$ on the combined fields whose interaction parameters are related by eq.~(\ref{OmegaDelatOmega2}).

The bottom panels of Fig.~\ref{fig:V_minus_plus_m01} show the 3D superpartner potentials, $V_\mp^\text{(3D)}(\theta)$, for a molecule in the combined fields along with the ground-state wavefunctions, $\psi_0(\theta)$, for $|m|=0,1$. The permanent dipole
term ($\propto \cos\theta$) is asymmetric with respect to $\theta =\pi /2$ whereas the
induced-dipole term ($\propto \cos^2\theta$) is symmetric about $\theta =\pi /2$ and gives rise to a double-well for $\alpha
_{\parallel }>\alpha _{\perp }$. At large $\beta$, $\omega \ll \Delta \omega $ and the coupling between the double-well's tunneling doublets results in localizing the ground-state wavefunction in the forward well of either the $V^{(1D)}$ or $V^{(3D)}$ potential, which  for $\omega \ll \Delta \omega $ is mainly due to the polarizability interaction. At small $\beta$, $\omega $ is comparable with $\Delta \omega $ and the effective potential becomes skewed as the forward well becomes deeper at the expense of the backward well. For any $\beta>0$, the molecule exhibits orientation (as opposed to alignment), which rapidly increases with $\beta$. 

We note that for $m=0$ the ground-state level always lies deeper than the potential minima at $\theta=\pi$, since $V(\theta= \pi) - E_0 \equiv 2 \beta >0$. Therefore, for any value of $\beta$ the molecule remains confined to the potential minimum at $\theta=0$ and exhibits orientation. For the ``stretched'' states of highly rotationally excited molecules, $J=m \gg 1$, the superymmetric partner potentials (\ref{Veff3D}) and (\ref{Vplus3D}) coincide.

\section{Shape invariance and exact solvability of the molecular Stark effect problem}
\label{sec:SIP}

Gendenshtein~\cite{Gendenshtein83} demonstrated that the Schr\"odinger equation for any of the superpartner potentials, $V_- (\theta, a)$ and $V_+ (\theta, a)$, admitting normalizable solutions, is exactly solvable if the potentials are translationally shape-invariant, i.e.

\begin{equation}
	\label{ShapeInv}
	V_+(\theta, a_0) + g(a_0) = V_- (\theta, a_1) + g(a_1),
\end{equation}
where the values $a_0$ and $a_1$ of the $a$ parameter pertain to the $n$-th eigenstate with $n=0$ and $n=1$, and the function $g(a)$ is independent of $\theta$. The parameter $a_1$ is a function of $a_0$, i.e.\ $a_1 = p(a_0)$.
 
If the superpartner potentials satisfy relation~(\ref{ShapeInv}), the eigenenergies and wavefunctions of $H_-$ can be obtained from~\cite{DuttKhareSukhatmeAJP88, GangopadhyayaIJMP08}:
\begin{equation}
	\label{ESIP}
	E_n^- = g\left( p^{(n)} (a_0) \right) - g (a_0),
\end{equation}
\begin{equation}
	\label{WFSIP}
	f_n^- (\theta, a_0) = A^\dagger (\theta, a_{n-1}) f_0^- (\theta, a_n),
\end{equation}
where $p^{(n)} (a_0) $ designates the function $p(a)$ applied $n$ times. 

Recently, Gangopadhyaya and Mallow demonstrated that
condition~(\ref{ShapeInv}) is equivalent to the following partial
differential equation for the superpotential~\cite{BougieMallowPRL10,
GangopadhyayaIJMP08}:
\begin{equation}
	\label{Wpartial}
	2 W(\theta, a) \left( \frac{\partial W(\theta, a)}{\partial a} \right )- 2 \frac{\partial W(\theta, a)}{\partial \theta} + \frac{\partial g(a)}{\partial a}  = 0
\end{equation}
Unfortunately, equation~(\ref{Wpartial}) is not satisfied for the superpotential $W(\theta, a)$ of eq.~(\ref{W}), neither for $a=\beta$, nor for $a=m$, and, therefore, the partner potentials $V_-^\text{(1D)} (\theta, a)$ and $V_+^\text{(1D)} (\theta, a)$ are not  shape invariant. Although, shape invariance is a sufficient, but not necessary condition for exact solvability~\cite{CooperPRD87}, Scr\"odinger equation for a molecule in combined fields is known to be in general unsolvable exactly~\cite{FriHerJCP99, AbramowitzStegun}. We also note that none of the known shape-invariant superpotentials listed, e.g., in refs.~\cite{DuttKhareSukhatmeAJP88, GangopadhyayaIJMP08}, leads to exactly solvable partner Hamiltonians that can be experimentally implemented for molecules in nonresonant fields. For reference purposes, we provide a summary of the relationships among exactly solvable, shape invariant, and Infeld-Hull factorizable potentials in Appendix~\ref{app}.

\section{Supersymmetry and shape invariance of the combined-fields Hamiltonian in the field-free and strong-field limits}
\label{sec:Limits}

%\begin{figure}[htbp]
%\includegraphics[width=6cm]{corr.pdf}
%\caption{\label{fig:corr} Energy levels of a molecule in combined fields for different values of $\beta$.  In the weak-field limit, $\beta \to 0$, the energy levels approach those of a free-rotor with a good quantum number $J\ge m$. For nonzero field strengths the levels exhibit Stark shift and are labeled by the adiabatic label $\tilde{J}$. In the strong-field limit, $\beta \to \infty$, the energy levels become equidistant and are labeled by the angular harmonic librator quantum number $n$. The exactly solvable cases are shown in blue. The energy spacings are represented only schematically for convenience.}
%\end{figure}

In the field-free limit, $\omega, \Delta \omega \to 0$, the 3D Schr\"odinger equation reduces to one for a rigid rotor:
\begin{equation}
	\label{SErotor3D}
	\mathbf{J}^2 \psi(\theta) = E\psi(\theta),
\end{equation}
whose solutions are spherical harmonics, $\psi(\theta) = Y_{J m} (\theta, 0)$ pertaining to eigenenergies $E_J = J(J+1)$.

\begin{figure}[htbp]
\includegraphics[width=12cm]{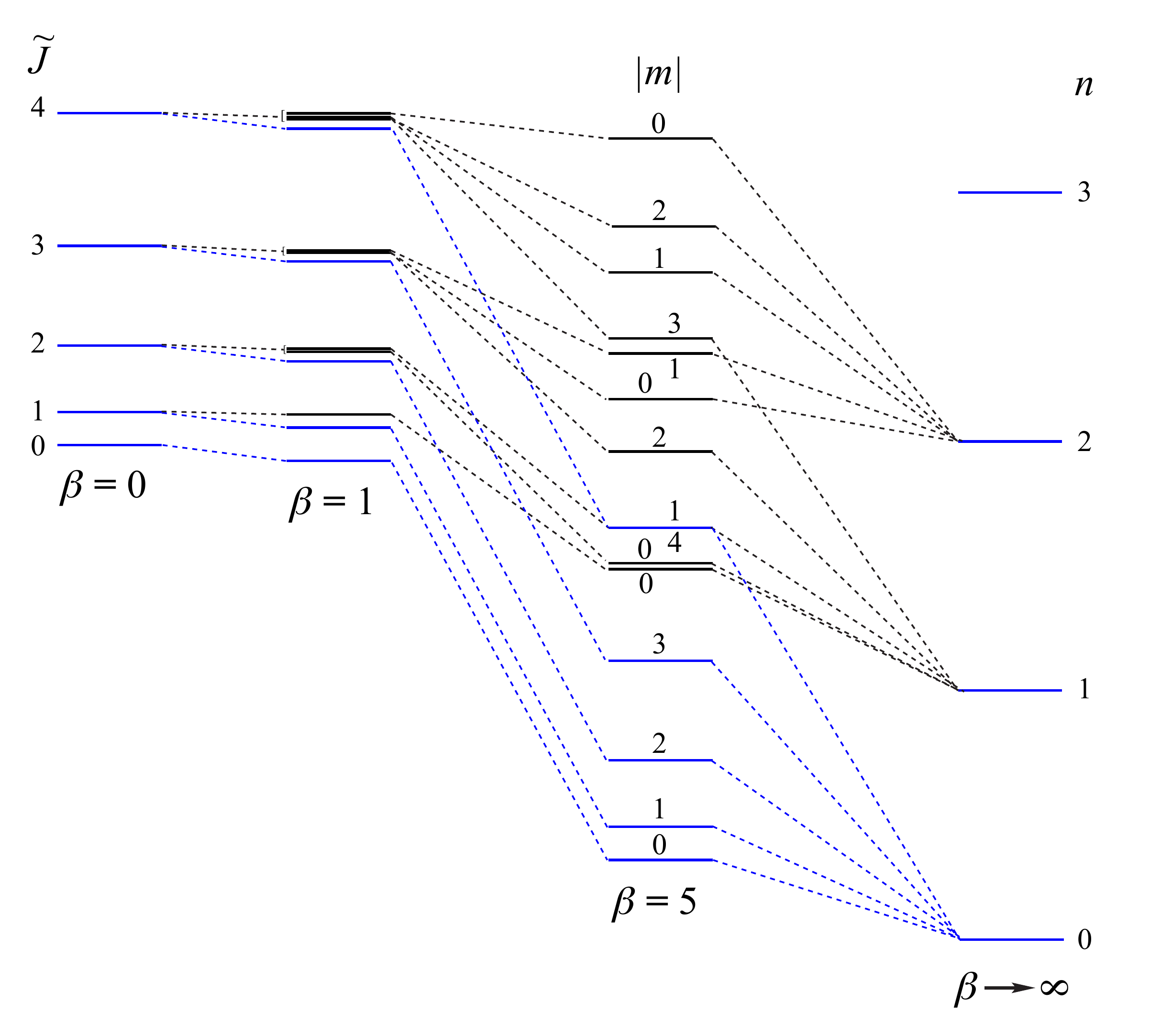}
\caption{\label{fig:corr} Eigenstates of a molecule in combined fields for different values of $\beta$,  with exactly solvable cases shown in blue. The equidistant energy levels in the strong-field limit, $\beta \to \infty$, are labeled by the 3D librator quantum number $n$ and are shown schematically. See text.}
\end{figure}

We obtain the expressions for the superpotential and partner potentials of a free rotor by setting $\beta=0$ in eqs.~(\ref{W}), (\ref{Vminus1D}), (\ref{Vplus1D}):
 \begin{equation}
	\label{WfreeRot}
	W(\theta) =  - \left( m + \frac{1}{2} \right) \cot \theta,
\end{equation}
\begin{equation}
	\label{Vminus1DRot}
	V_-^\text{(1D)}(\theta) =   \frac{m^2 - \frac{1}{4}}{\sin^2 \theta} -  m(m+1) -\frac{1}{4},
\end{equation}
and
\begin{equation}
	\label{Vplus1DRot}
	V_+^\text{(1D)}(\theta) =    \frac{(m+1)^2 - \frac{1}{4}}{\sin^2 \theta} -  m(m+1) -\frac{1}{4}.
\end{equation}
This is an obvious result as for $J=m$, the spectra of two rigid rotors with quantum numbers $m$ and $m+1$ coincide except for the ground state~\cite{InfeldHullRMP51, DuttAJP97}.

Potentials (\ref{Vminus1DRot}) and (\ref{Vplus1DRot}) are seen to be shape-invariant, with  $a_0 = m$, $a_1\equiv p(a_0) = a_0+1$, and $g(a_n) = a_n(a_n+1)$, and the superpotential of eq. (\ref{WfreeRot}) satisfies eq.~(\ref{Wpartial}). The eigenenergies of the Hamiltonian $H_-$ are obtained from eq.~(\ref{ESIP}):
\begin{equation}
	\label{EminusFreeRotor}
	E_n^- = n (n + 2m + 1),
\end{equation}
which for $n=J-m$ coincides with the rigid rotor spectrum, shifted by $-m(m+1)$. The energy spectrum and wavefunctions of the supersymmetric partner potential $V_+$ can be obtained via eqs.~(\ref{Erelation}) and (\ref{PsiplusiaPsiminus}).

In the strong-field limit, $\omega, \Delta \omega \to \infty$, the angular motion of the molecules is confined near $\theta=0$ and potential~(\ref{Veff1D}) can be expanded in powers of $\theta$. On retaining terms up to second order, the potential can be reduced to that of a 3D angular oscillator (librator). The corresponding Schr\"odinger equation~(\ref{SEtransformed}) takes the form:
\begin{equation}
	\label{SE3Dlibrator}
	\left [ -\frac{d^2}{d\theta^2} + \frac{\kappa^2 \theta^2}{4} + \frac{\ell(\ell+1)}{\theta^2} \right ]f(\theta) = \tilde{E} f(\theta),
\end{equation}
where the eigenenergies are given by
\begin{equation}
	\label{EnergOscill}
	\tilde{E}_n \equiv E + \omega + \Delta \omega + \frac{1}{4}
= \kappa \left (2 n + \ell + \frac{3}{2} \right), \hspace{0.4cm} n =0, 1, 2, \dots
\end{equation}
and $\kappa^2 = 2 (\omega + 2\Delta \omega)$, $\ell = \left (\pm m-1/2 \right)$. 

The analytic eigenfunctions of the angular harmonic oscillator can be expressed in terms of Laguerre polynomials~\cite{CooperPhysRep95},
\begin{equation}
	\label{RadOscillWF}
	f_n (\theta) = N \theta^{\ell+1} L_n^{(\ell+1/2)}(x) \exp(-x/2), \hspace{0.4cm} x \equiv \tfrac{1}{2} \kappa \theta^2,
\end{equation}
with
\begin{equation}
	\label{RadOscNorm}
	N = \left( \frac{\kappa}{2} \right)^{\ell/2+3/4} \left[ \frac{2n!}{\Gamma (n+\ell +3/2) } \right]^{1/2}
\end{equation}
and they vanish at $x \to 0$ and decay exponentially at $x \to \infty$.

The corresponding superpotential can be constructed from the ground state wavefunction, eq.~(\ref{RadOscillWF}), with $n=0$, 
\begin{equation}
	\label{RadSocW}
	W(\theta) \equiv - \frac{f_0'(\theta)}{f_0(\theta)} = \frac{\kappa \theta}{2} - \frac{\ell+1}{\theta},
\end{equation}
which leads to the supersymmetric partner potentials
\begin{equation}
	\label{VminusRadSoc}
	V_- (\theta) = \frac{\kappa^2 \theta^2}{4} + \frac{\ell(\ell+1)}{\theta^2} - \kappa \left( \ell + \frac{3}{2} \right) 
\end{equation}
\begin{equation}
	\label{VplusRadSoc}
	V_+ (\theta) = \frac{\kappa^2 \theta^2}{4} + \frac{(\ell+1)(\ell+2)}{\theta^2} - \kappa \left( \ell + \frac{1}{2} \right) 
\end{equation}
The potential of eq. (\ref{VminusRadSoc}) coincides with the potential of eq.~(\ref{SE3Dlibrator}) shifted by the ground state energy, $E_0 = - \kappa \left( \ell + 3/2 \right) $. 

Potentials (\ref{VminusRadSoc}) and (\ref{VplusRadSoc}) are shape-invariant, with the parameters $a_0 =  \ell$, $a_1\equiv p(a_0) = a_0+1$, and $g(a_n) =2 \kappa a_n$ and superpotential (\ref{RadSocW}) satisfies eq.~(\ref{Wpartial}). The eigenenergies of the Hamiltonian $H_-$ can be obtained from eq.~(\ref{ESIP}):
\begin{equation}
	\label{EminusFreeRotor}
	E_n^- = 2 \kappa n,
\end{equation}
which coincides with the eigenenergies given by eq.~(\ref{EnergOscill}) shifted by $- \kappa \ell - 3/2$. The eigenenergies and eigenfunctions of the superpartner potentials also fulfill eqs.~(\ref{Erelation})--(\ref{gr}) and represent a complete analytic solution to the problem.

Figure~\ref{fig:corr} shows the energy levels of a molecule in combined fields for different values of the field-strength parameter $\beta$. In the weak-field limit, $\beta \to 0$, the energy levels approach those of a free-rotor, which is solvable exactly for all the eigenstates. For nonzero but weak fields, $\beta=1$, the levels become split in $\tilde{J}+1$ components due to the Stark effect. In this case, the SUSY partner Hamiltonians are not shape-invariant, and the problem is analytically solvable only for the ``stretched states,'' corresponding to $\tilde{J} = m$. With increasing interaction strength, the energies of the stretched states come closer to one another and, in the strong-field limit, $\beta \to \infty$, coalesce into the ground state level of the 3D harmonic librator. In the strong-field limit, the supersymmetric problem becomes shape-invariant again, and is exactly solvable for all eigenstates in closed form; the equidistant levels are infinitely degenerate and separated by an energy difference of $(2\omega + 4 \Delta \omega)^{1/2}$.

\section{Applications}
\label{sec:Applications}

\begin{figure}
\includegraphics[width=7.2cm]{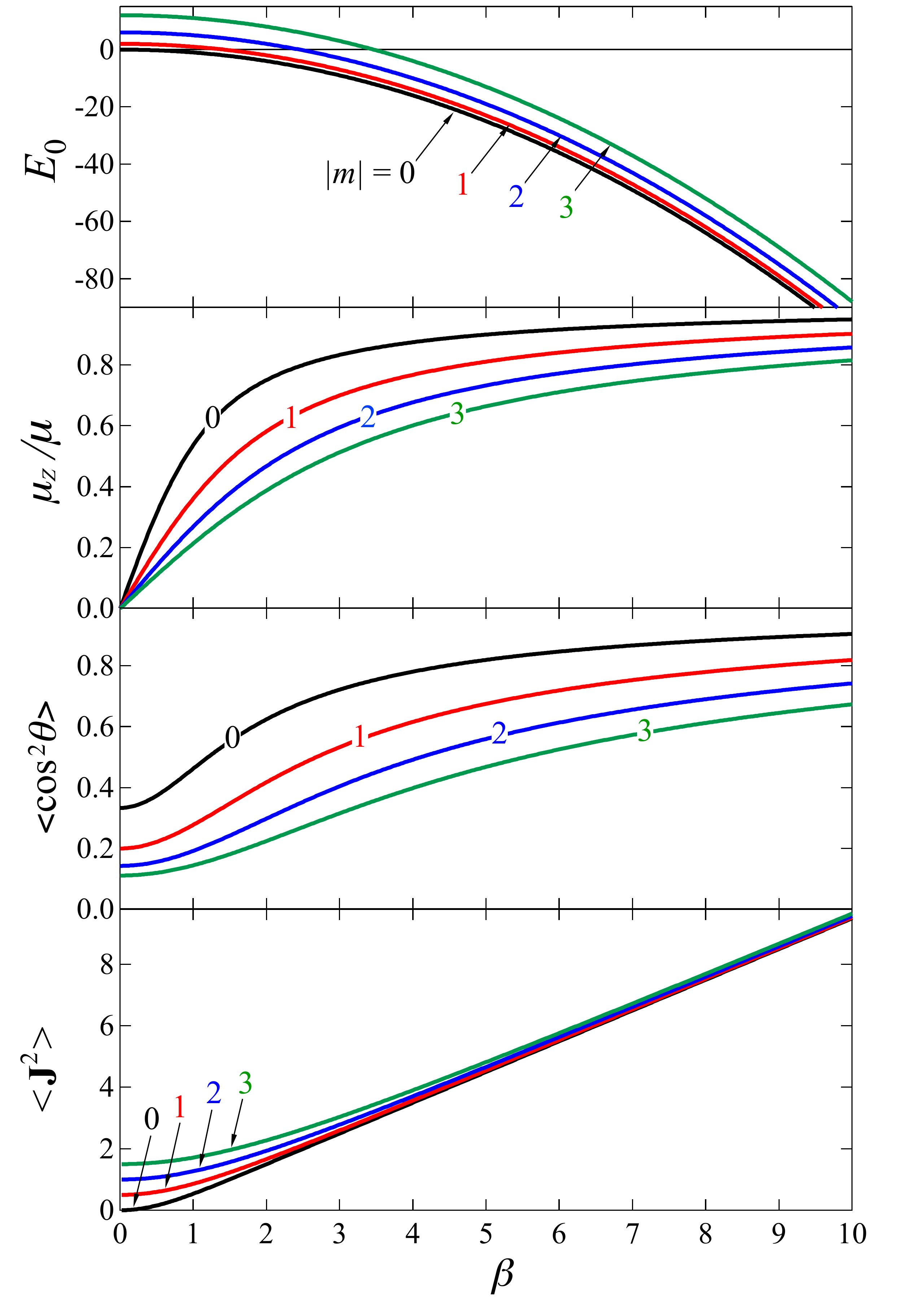}
\caption{\label{fig:mu_Z} Ground-state energies $E_0$ (in units of the rotational constant $B$), space fixed dipole moments $\mu_Z/\mu \equiv \langle \cos \theta \rangle$, alignment cosines $\langle \cos^2 \theta \rangle$, and expectation values of the angular momentum $\langle \mathbf{J}^2 \rangle$ for different $|\tilde{J}=m,m;\beta \rangle$ states as a function of the interaction parameter $\beta$.}
\end{figure}

The analytic wavefunctions of molecules in combined fields make it possible to obtain their properties analytically as well. The space fixed dipole moment, $\mu_Z$, is given by the orientation cosine, $\langle \cos \theta \rangle = \langle \psi (\theta) \vert \cos \theta \vert \psi (\theta) \rangle $, and for the exact wavefunction of eq.~(\ref{psiWF}) can be evaluated in closed form,
\begin{equation}
	\label{DipoleMom}
	\mu_Z / \mu \equiv  \langle \cos \theta \rangle  = \frac{ I_{m+3/2} (2 \beta)} {I_{m+1/2} (2 \beta)},
\end{equation}
where $I_n (z)$ is a modified Bessel function of the first kind~\cite{AbramowitzStegun}. 
Fig.~\ref{fig:mu_Z} shows spaced fixed dipole moments corresponding to the states $|\tilde{J}=m,m;\beta \rangle$ for several values of  $m$ as a function of the $\beta$ parameter. The value of $\mu_Z$ rapidly increases with $\beta$. For instance, for $m=0$,  it rises from only $0.54 \mu$ at $\beta=1$ to $0.83 \mu$ at $\beta=3$. In the case of the much  studied $^{40}$K$^{87}$Rb molecule, which possesses a dipole moment $\mu=0.589$ Debye and a polarizability anisotropy  $\Delta \alpha = 54.21$~\AA$^3$~\cite{AymarDulieuJCP05, DeiglmayrDulieuJCP08}, relatively weak fields of $\varepsilon=38$ kV/cm and $I=1.75 \cdot 10^9$ W/cm$^2$ (corresponding to $\beta=5$) give rise to a strongly oriented ground state with $\mu_Z = 0.9 \mu$. This value of $\langle \cos \theta \rangle$ corresponds to the molecular axis confined to librate within $\pm26^\circ$ about the common direction of the fields.

The alignment cosine, $\langle \cos^2 \theta \rangle = \langle \psi (\theta) \vert \cos^2 \theta \vert \psi (\theta) \rangle $, characterizes the molecular alignment along the $Z$ axis. For the wavefunction of eq.~(\ref{psiWF}), the alignment cosine takes the analytic form,
\begin{equation}
	\label{cos2}
	\langle \cos^2 \theta \rangle  = \frac{2\beta^2 {}_0\tilde{F}_1 (; m+7/2; \beta^2) + {}_0\tilde{F}_1 (; m+5/2; \beta^2) }{2 {}_0\tilde{F}_1 (; m+3/2; \beta^2)},
\end{equation}
with ${}_0\tilde{F}_1 (; a; z) = {}_0F_1(; a; z)/\Gamma(a)$ a regularized confluent hypergeometric function~\cite{AbramowitzStegun}. Fig.~\ref{fig:mu_Z} shows $\langle \cos^2 \theta \rangle$  of the states $|\tilde{J}=m,m;\beta \rangle$ for several values of  $m$ as a function of the $\beta$ parameter. The $|0,0;\beta \rangle$ state exhibits quite a strong alignment with the alignment cosine rapidly approaching with increasing $\beta$ the value of 0.8, which corresponds to a libration of the molecular axis about the polarization vector of the radiative field with an angular amplitude of $27^\circ$.

The expectation value of the angular momentum is related to the orientation cosine, eq.~(\ref{DipoleMom}), via
\begin{equation}
	\label{J2}
	 \langle  \mathbf{J}^2  \rangle  = \frac{m}{2} + \beta \frac{I_{m+3/2} (2 \beta)} {I_{m+1/2} (2 \beta)} \equiv \frac{m}{2} + \beta \langle \cos \theta \rangle,
\end{equation}
We note that the dependence of $\langle  \mathbf{J}^2  \rangle$, shown in Fig.~\ref{fig:mu_Z}, becomes asymptotically linear in $\beta$ for all the values $|m|$, cf. eq.~(\ref{J2}).

By making use of eqs.~(\ref{SEcombinedfields}) and (\ref{OmegaDeltaOmega}), one can show that the following condition for the expectation values is satisfied:
\begin{equation}
	\label{ExpectValuesCond}
	 \langle  \mathbf{J}^2  \rangle + m^2 \biggl < \frac{1}{\sin^2\theta} \biggr > - 2\beta (m+1) \langle \cos \theta \rangle - \beta^2 \langle \cos^2 \theta \rangle \equiv E_0,
\end{equation}
with the energy $E_0$ given by eq.~(\ref{EnergPsi}).

Table~\ref{table:muZ} lists, for reference purposes, the above analytic expressions for the space-fixed dipole moment, alignment cosine, and expectation value of the angular momentum for different values of $|m|$.

\begin{table}[h]
\centering
\caption{Analytic expressions for the space-fixed dipole moment $\mu_Z/\mu \equiv \langle \cos \theta \rangle$, alignment cosine $\langle \cos^2 \theta \rangle$, and expectation value of the angular momentum $\langle \mathbf{J}^2 \rangle$ obtained from eqs.~(\ref{DipoleMom}), (\ref{cos2}), and (\ref{J2}) for different values of $|m|$.\\}
\label{table:muZ}
\begin{tiny}
\begin{tabular}{| c | c | c | c |}
\hline 
\hline
$|m|$ & $\mu_Z / \mu \equiv \langle \cos \theta \rangle $ & $\langle \cos^2 \theta \rangle$ & $\langle \mathbf{J}^2 \rangle$  \\[3pt]
\hline
 0 &  $\coth(2\beta) - \dfrac{1}{2\beta}$   &   $1+ \dfrac{1}{2\beta^2} - \dfrac{\coth(2\beta)}{\beta}$    & $\beta \coth(2\beta) - \dfrac{1}{2}$  \\[10pt]
 1 & $\dfrac{2 \beta}{2\beta \coth(2\beta)-1} - \dfrac{3}{2\beta}$     &  $5+ \dfrac{3}{\beta^2} + \dfrac{8 \beta}{\tanh(2\beta) - 2\beta}$  &  $\dfrac{2\beta^2}{2\beta\coth(2\beta) -1}  - 1$     \\[15pt]
 2 &  $\dfrac{8 \beta^3}{3 \left( 4\beta^2 - 6\beta \coth(2\beta) + 3  \right)} - \dfrac{2\beta}{3} - \dfrac{5}{2\beta}$      &   $3+ \dfrac{15}{2 \beta^2} - \dfrac{8\beta^2}{3+ 4\beta^2 - 6\beta \coth(2\beta)}$   
 & $- \dfrac{3}{2} - \dfrac{2\beta^2}{3} + \dfrac{8\beta^4}{3(3+4\beta^2 - 6\beta \coth(2\beta))}$   \\[15pt]
 3 & $\dfrac{(16 \beta^4 + 180 \beta^2 +105) - 10\beta (8\beta^2 + 21) \coth(2\beta)}{4 \beta^2 (4\beta^2 +15) \coth(2\beta) - 6\beta (8 \beta^2 +5)}$    &   $\dfrac{2\beta (4\beta^4 + 95 \beta^2 + 210) \coth(2\beta) - (56\beta^4 + 375\beta^2 + 210)}{2\beta^3 (4\beta^2 + 15)\coth(2\beta) - 3\beta^2 (8\beta^2 +5)}$  &   $\dfrac{-4\beta (7\beta^2 +15) \coth(2\beta) +2 (4\beta^4 + 27 \beta^2 + 15) }{2 \beta (4\beta^2+15) \coth(2\beta) - 3(8\beta^2 +5)} $  \\[10pt]
 \hline
 \hline
\end{tabular}
 \end{tiny}
\end{table}

\section{Conclusions}
\label{sec:Conclusions}

By invoking supersymmetry, we found a condition under which the molecular Stark effect problem becomes exactly solvable. The condition, $\Delta \omega = \frac{\omega^2}{4 (m+1)^2 }$, cf. eq.~(\ref{OmegaDeltaOmega}),  connects values of the parameters $\omega$ and $\Delta \omega$ that characterize the interaction strengths of a polar and polarizable molecule with collinear electrostatic and nonresonant radiative fields. The exact solutions are obtained for the  $|\tilde{J}=m,m;\omega,\Delta \omega\rangle$ family of ``stretched'' states.

We also considered the field-free and strong-field limits of the combined-fields problem and found that both  exhibit supersymmetry and shape-invariance, which is indeed the reason why they are analytically solvable. 

By making use of the analytic form of the $|\tilde{J}=m,m;\omega,\Delta \omega\rangle$  wavefunctions, we derived simple analytic  formulae for the expectation values of the space-fixed electric dipole moment, the alignment cosine, and the angular momentum. These key characteristics of molecules in fields are summarized in Table~\ref{table:muZ}. We also derived a ``sum rule'' which yields a formula for the corresponding eigenenergy, eq. (\ref{ExpectValuesCond}), in terms of the above expectation values. The analytic expressions obtained open a direct route to engineering molecular states with preordained characteristics.

Interestingly, it is possible to glean the reason as to why the exact solution of eq. (\ref{SEtransformed}) is obtained for only one particular relation between the field strength parameters, eq. (\ref{OmegaDeltaOmega}), from the semiclassical (WKB) approximation~\cite{FriedrichTrostPhysRep04}. The eigenfunction of a 1D Schr\"odinger equation assumes the WKB form $f(\theta) \propto \exp [i S(\theta)]$, with $S(\theta)$ the action of the underlying classical system. For the ground state of the potential~(\ref{Vminus1D}), the action satisfies the differential equation, $S'(\theta)^2 - i S''(\theta) = V_-^\text{(1D)}$, whose solutions are obtained by expanding $S(\theta)$ in powers of $\hbar$. It turns out that in the case of the combined field strengths connected via eq.~(\ref{OmegaDeltaOmega}), the series converges to the following exact expression,
\begin{equation}
	\label{Action}
	S (\theta) =  \frac{1}{2 i} \left[ 2 \beta \cos \theta + (2m+1) \ln (\sin \theta) \right],
\end{equation}
which, when substituted into $f(\theta)$, yields the exact, closed form wavefunction, eq. (\ref{fviaW}).
  
We note that the exact ground-state wavefunction $|\tilde{J}=0,m=0;\omega=0,\Delta \omega\rangle$ can be also obtained as a ``curious eigenproperty'' by the method outlined by von Neumann and Wigner in 1929. They showed that by imposing the integrability condition on the sought wavefunction, a class of potentials could be derived that support a localized bound state embedded in the continuum ~\cite{vonNeumannWigner29, StillingerHerrickPRA75, MeyerVernet82}.

An extension of the method described in Section~\ref{sec:SUSYfactor} might furnish other types of exact wavefunctions for the case of more than two combined fields and for non-collinear field geometries.

\section{Acknowledgements}

Our special thanks are due to Gerard Meijer for encouragement and support. One of us (S.K.) would like to thank the ARO for financial support.

\appendix

\section{Exactly solvable, shape invariant, and Infeld-Hull factorizable
potentials}
\label{app}

\begin{figure}[htp]
\centering
    \includegraphics[width=.3 \textwidth]{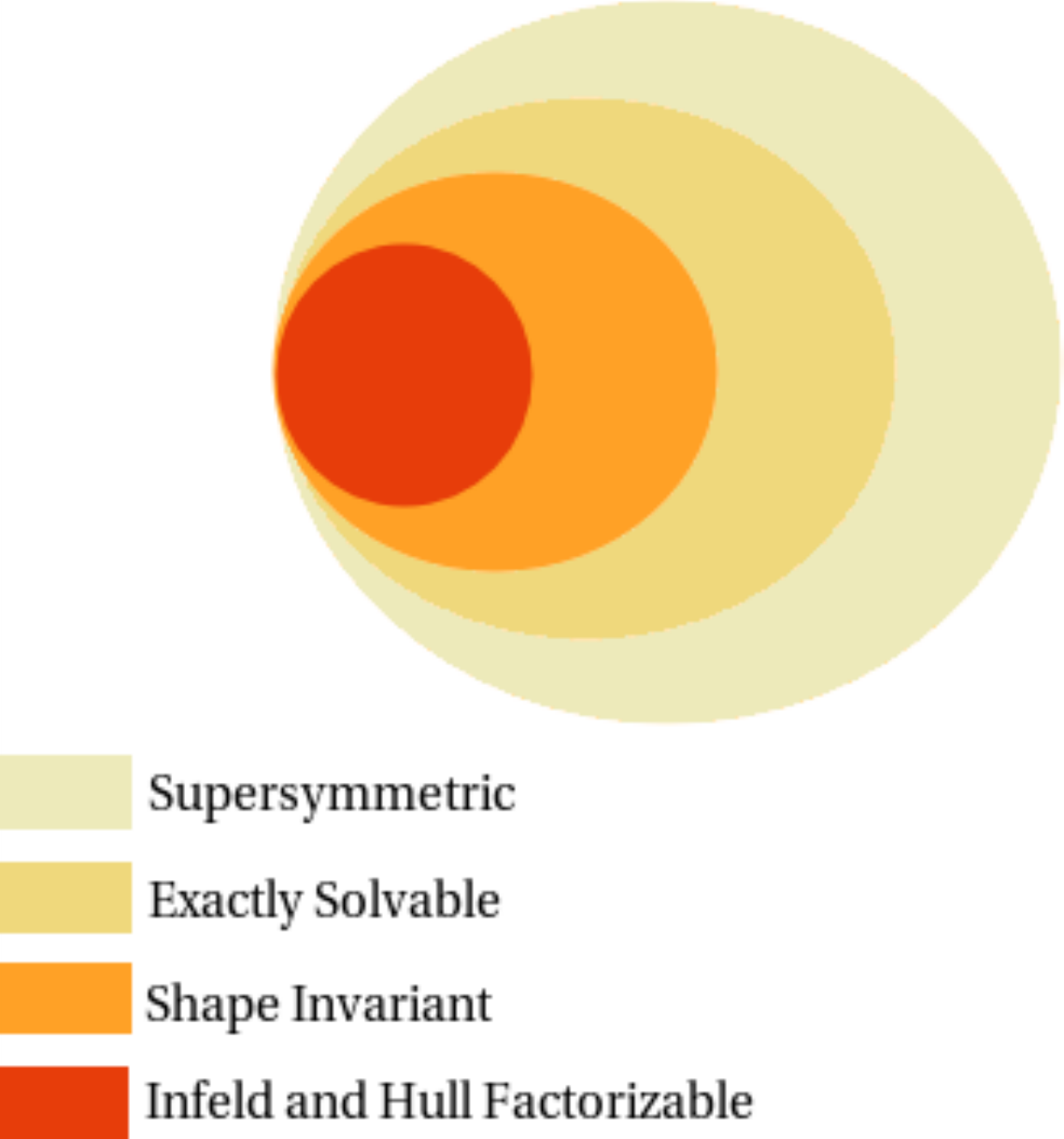}
  \caption{A Venn diagram showing the relationhips among supersymmetric, exactly
solvable, shape invariant, and Infeld \& Hull factorizable potentials.}
\label{venn}
\end{figure}

Supersymmetry, shape invariance, exact solvability, and the factorization method
are often studied concurrently in the literature. In this appendix we briefly survey the literature studies on the relationships among them. Figure \ref{venn} summarizes the results.
\newline\newline
\textit{Every one-dimensional potential with a ground state in close form admits SUSY.} Given a
one dimensional quantum system with at least one bound state, one can find
a partner Hamiltonian which has exactly the same discrete spectrum except for
the ground state energy of $H_-$~\cite{Sukumar19852917, Montemayor1989}.
\newline\newline
\textit{Every Infeld and Hull factorizable potential is shape invariant but the
converse is not true.} Shape invariance offers, in general, more than the
factorization method, since the factorization method treats only the
translational shape invariance~\cite{Montemayor1989,Carinena2000}. The table
prepared by Infeld and Hull~\cite{InfeldHullRMP51}, although complete for most purposes, is not
the most general table possible, as it does not include the most general
solution of the Ricatti equations,  worked out in Ref. \cite{Carinena2000}.
\newline\newline
\textit{Shape invariance and normalizability are sufficient but not necessary conditions for exact
solvability}. Gendenshtein suggested that all exactly solvable potentials must
be shape invariant~\cite{Gendenshtein1983}, but many counter examples to this conjecture were later constructed, for instance the Natanzon class of potentials which are, in general,
not shape invariant~\cite{Cao1991,CooperPRD87,cooper1989}. We also note that in order to be exactly solvable, the shape-invariant potentials should admit normalizable solutions.
\newline\newline

\bibliography{References_library}
\end{document}